\documentclass[iop]{emulateapj}

\begin{document}

\title{{\it Chandra} Reveals Heavy Obscuration and Circumnuclear Star Formation in Seyfert 2 Galaxy NGC 4968}

\author{Stephanie M. LaMassa$^{1,2}$, Tahir Yaqoob$^3$, N. A. Levenson$^4$, Peter Boorman$^5$, Timothy M. Heckman$^6$, Poshak Gandhi$^5$, Jane R. Rigby$^2$, C. Megan Urry$^{7}$, Andrew F. Ptak$^2$ }
\affil{$^1$NASA Postdoctoral Program Fellow;
  $^2$NASA Goddard Space Flight Center, Greenbelt, MD 20771, USA;
  $^3$Department of Physics, University of Maryland Baltimore County, 1000 Hilltop Circle, Baltimore, MD 21250, USA;
  $^4$Gemini Observatory, Casilla 603, La Serena, Chile;
  $^5$Department of Physics \& Astronomy, University of Southampton, Southampton, SO17 1BJ, UK;
  $^6$Department of Physics \& Astronomy, The Johns Hopkins University, 3400 North Chales Street, Baltimore, MD 21218, USA;
  $^7$Yale Center for Astronomy \& Astrophysics, Physics Department, P.O. Box 208120, New Haven, CT 06520, USA;
}
\begin{abstract}
  We present the {\it Chandra} imaging and spectral analysis of NGC 4968, a nearby ($z = 0.00986$) Seyfert 2 galaxy. We discover extended ($\sim$1 kpc) X-ray emission in the soft band (0.5 - 2 keV) that is neither coincident with the narrow line region nor the extended radio emission. Based on spectral modeling, it is linked to on-going star formation ($\sim$2.6-4 M$_{\sun}$ yr$^{-1}$). The soft emission at circumnuclear scales (inner $\sim$400 pc) originates from hot gas, with kT $\sim$ 0.7 keV, while the most extended thermal emission is cooler (kT $\sim$ 0.3 keV). We refine previous measurements of the extreme Fe K$\alpha$ equivalent width in this source (EW =  2.5$^{+2.6}_{-1.0}$ keV), which suggests the central engine is completely embedded within Compton-thick levels of obscuration. Using physically motivated models fit to the {\it Chandra} spectrum, we derive a Compton-thick column density ($N_{\rm H} > 1.25\times10^{24}$ cm$^{-2}$) and an intrinsic hard (2-10 keV) X-ray luminosity of  $\sim$3-8$\times 10^{42}$ erg s$^{-1}$ (depending on the presumed geometry of the obscurer), which is over two orders of magnitude larger than that observed. The large Fe K$\alpha$ EW suggests a spherical covering geometry, which could be confirmed with X-ray measurements above 10 keV. NGC 4968 is similar to other active galaxies that exhibit extreme Fe K$\alpha$ EWs (i.e., $>$2 keV) in that they also contain ongoing star formation. This work supports the idea that gas associated with nuclear star formation may increase the covering factor of the enshrouding gas and play a role in obscuring AGN.

\end{abstract}

\section{Introduction}
Active galactic nuclei (AGN) are powered by accreting supermassive black holes in galactic centers. A significant portion of this black hole growth is obscured by gas and dust \citep[e.g.,][]{risaliti99,fabian99,treister04,treister}, which according to the AGN unification model, resides on $\sim$1-10 pc scales in a toroidal geometry \citep{antonucci,urry}. According to this model, Type 2 AGN lack broad lines in their optical spectra due to the observer line of sight intersecting the obscuring torus, while Type 1 AGN are unobscured, allowing both broad emission lines and the AGN continuum to be visible.

Compton-thick AGN, in which the column density ($N_{\rm H}$) of the obscuring material becomes optically thick to Compton scattering ($\geq$1.25 $\times$ 10$^{24}$ cm$^{-2}$),\footnote{Accounting for the contribution of electrons from Hydrogen and Helium, the mean number of electrons per Hydrogen atom is 1.2 so that the Compton optical depth is N$_{\rm H} = (1.2\times \sigma_{\rm T})^{-1}$, where $\sigma_{\rm T}$ is the Thomson cross-section \citep[see, e.g.,][]{mytorus}} appear to represent a significant fraction of cosmic black hole growth, though the percentage of this population is a matter of some debate \citep[e.g.,][]{treister05,gilli,treister,ueda,buchner,ricci15,aird,akylas}. Hence, to constrain population synthesis models and accurately determine the accretion history of cosmic black hole growth, it is important to have a proper census of Compton-thick AGN. These heavily obscured sources can be identified by signatures in their X-ray spectra, with the fluorescent Fe K$\alpha$ line emission at 6.4 keV being perhaps the most notable feature \citep{krolik,ghisellini,matt}. Detailed study of individual Compton-thick AGN then provides an opportunity to probe AGN physics. Furthermore, the suppression of optical AGN emission allows us to study the host galaxy in greater detail than is possible with unobscured AGN since galaxy features are not swamped by AGN light. Circumnuclear star formation may even play a role in obscuring some AGN \citep[e.g.,][]{ballantyne}. 

NGC 4968 is a nearby ($z=0.00986$) Type 2 Seyfert (Sy2) identified in the {\it IRAS} 12$\mu$m sample \citep{12um}, hosted in an Sa galaxy \citep{malkan} with a {\it WISE} measured 12$\mu$m luminosity of 1.7$\times10^{43}$ erg s$^{-1}$ \citep{wright}. Our X-ray analysis of this source with {\it XMM-Newton} revealed a strong Fe K$\alpha$ line with equivalent width (EW) of 3.10$^{+1.0}_{-0.8}$ keV (\citealp{me2011}; see also \citealp{guainazzi}). This extreme value indicates that the AGN in NGC 4968 is embedded within Compton-thick material: the total continuum from the transmitted and reflected components is suppressed relative to the fluorescent line photons which can be created anywhere throughout the obscuring medium. However, using a phenomenological model of an absorbed power law plus a Gaussian component to fit the AGN continuum and Fe K$\alpha$ line, respectively, we did not find an absorbing column larger than that from our Galaxy.  Excess emission at soft energies (0.5-2 keV) prevented the column density from being accurately measured with this overly simplistic model. Is this soft X-ray emission due to the host galaxy, scattered AGN emission, or both? \citet{levenson} noted that Sy2s with large Fe K$\alpha$ EW values ($>$2 keV), like NGC 4968, reside in galaxies with on-going star formation, so it is likely that star formation contributes to the soft X-ray emission in NGC 4968.

With the quality and resolution of the {\it XMM-Newton} data, we were unable to fit self-consistent, physically motivated models (e.g., MYTorus, \citealp{mytorus}; spherical and toroidal absorption models of \citealp{brightman}) to the spectra, nor could we attempt to disentangle nuclear from unresolved galaxy light. Therefore, we obtained a deep (50 ks) {\it Chandra} observation in Cycle 16 in order to physically diagnose the conditions at the center of this galaxy. We present our results in this paper as follows: in Section \ref{da}, we detail the reduction of the {\it Chandra} data. We discuss the imaging and spectral analysis of NGC 4968 in Sections \ref{imaging} and \ref{spec}, respectively. From these results, we comment on the nature of the X-ray obscuring medium, estimate the star formation rate of the host galaxy, and propose that the gas responsible for circumnuclear star formation increases the covering fraction of the obscuration towards unity, such that the AGN is almost completely enshrouded within Compton-thick material (Section \ref{disc}). Throughout, we use a cosmology where H$_{0}$=67.8 km s$^{-1}$ Mpc$^{-1}$, $\Omega_{M}$=0.31, $\Omega_{\Lambda}$=0.69 \citep{planck}.

\section{Data Analysis}\label{da}

NGC 4968 was observed for 50 ks on 2015 March 9 with ACIS-S (ObsID: 17126; PI: LaMassa). The data were reduced with \textsc{CIAO v4.8}, with \textsc{CALDB v. 4.7.1}. We used \textsc{CIAO} task \textsc{chandra\_repro} to produce a filtered level 2 events file with the latest calibration files.

\begin{figure}
  \centering
  \includegraphics[scale=0.25]{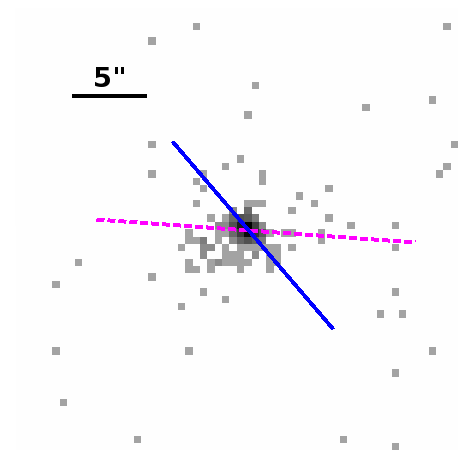}~
  \includegraphics[scale=0.25]{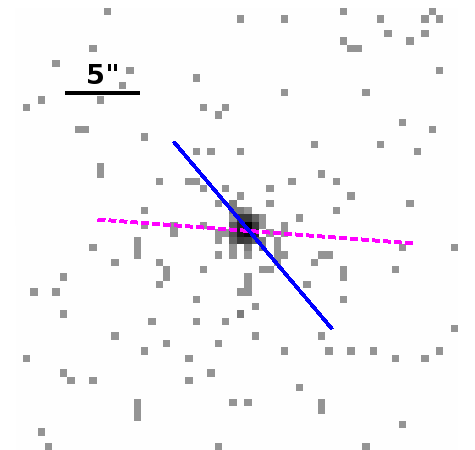}
  \includegraphics[scale=0.5]{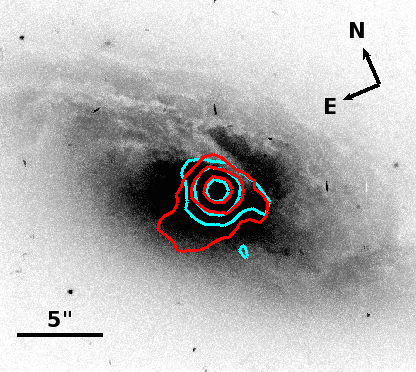}
  \caption{\label{image}{\it Top}: {\it Chandra} images of NGC 4968 in the soft (0.5-2 keV, {\it left}) and hard (2-10 keV, {\it right}) bands. The blue line denotes the position angle of the narrow line region, as traced by the [OIII] emission \citep{schmitt03}, and the dashed magenta line indicates the position angle of the radio emission \citep{schmitt01}; both emitting regions are smaller than indicated by the lines, extending 2.3$^{\prime\prime}$ and 2.2$^{\prime\prime}$, respectively. {\it Bottom:} {\it Hubble} WFPC2 image of NGC 4968 \citep{malkan} in the F606W filter with the X-ray contours in the soft (red) and hard (cyan) bands overplotted; the contours correspond to count levels of 2.8, 16.1, and 90.9 (soft band) and 1.8, 10.6, and 59.9 (hard band). The X-ray emission is spatially extended in the soft band (see Figure \ref{extent}), though confined to the bulge of the galaxy (in projection).  }
\end{figure}

From this events file, we extracted images of NGC 4968 in the soft (0.5-2 keV) and hard (2-10 keV) bands, as shown in Figure \ref{image} (top), where the location of the narrow line region \citep{schmitt03}, and slightly elongated radio emission \citep{schmitt01} are noted by the solid  blue and dashed magenta lines, respectively.\footnote{The lengths of the lines are larger than the size of the emitting regions to ease visualization.} As can be seen via visual inspection of the images, the source is more spatially extended in the soft band compared with the hard band (see Section \ref{imaging} for a full treatment of the spatial analysis). For reference, we show the {\it Hubble Space Telescope} WFPC2 image of NGC 4968 \citep{malkan} in the F606W filter in the bottom panel of Figure \ref{image}, overplotted with X-ray contours from the soft (red) and hard (cyan) bands. While the emission at softer energies has a larger spatial extent, it is confined to the bulge of the host galaxy (at least in projection).

We used \textsc{CIAO} routine \textsc{specextract} to extract the spectrum from the full X-ray emitting region as well as the nuclear and extended components separately. For the global modelling, the spectrum was extracted from a circular aperture with radius 6$^{\prime\prime}$ to ensure all extended emission was captured. To isolate the nuclear emission, we ran \textsc{wavdetect} on the hard energy band, extracting the spectrum from the output region file centered on the AGN (an ellipse with major axis 1.86$^{\prime\prime}$, minor axis 1.6$^{\prime\prime}$ and angle 137.4$^{\circ}$, measured counter-clockwise from West to East in WCS coordinates). This nuclear region was excised from the global region to produce a spectrum of the extended-only emission (as noted in Section \ref{imaging} the estimated PSF at the source location is smaller than the nuclear region). In all cases, the background was extracted from an annulus centered on the source with an inner radius of 10$^{\prime\prime}$ and outer radius of 30$^{\prime\prime}$.
The spectra were grouped by 5 counts per bin; visual inspection of the spectra demonstrates that the chosen binning does not over sample the resolution and thus we lose no information with this binning.

\section{Chandra Imaging Analysis}\label{imaging}
From Figure \ref{image}, the X-ray emission appears more extended in the soft band than in the hard band, and lies in a direction not coincident with the narrow line region \citep{schmitt03} nor radio structure \citep{schmitt01}. To test whether the X-ray emission is extended and to quantify the size of this region, we simulated a point spread function (PSF) at the position where the source was detected. Since the {\it Chandra} PSF varies as a function of energy as well as position, we fit the {\it Chandra} spectrum with an {\it ad hoc} model that reproduces the shape of the observed spectrum so that the energy dependence of the source can be accounted for by \textsc{ChaRT}, the {\it Chandra} ray tracing software (see Section \ref{spec} for a full treatment of the spectral modeling). The output ray-traced data from the simulated PSF are then input into \textsc{MARX} which projects the photons onto the ACIS-S detector-plane to create a pseudo-events file. We simulated pseudo-events files for a point source separately for the soft and hard band.

\textsc{CIAO} tool \textsc{srcextent} compares the PSF with the actual data to measure the size of the PSF at the source location on the detector and to quanity whether the source is extended at the 90\% confidence level. From this routine, we find that in both the soft and hard bands, the source is extended at the 90\% confidence level. The observed size of the PSF at the location of the source is 0$\farcs$69, where the PSF is measured as $\sqrt{(1/2)(a^2 + b^2)}$ with $a$ and $b$ as the semi-major and semi-minor axes of an elliptical Gaussian.

To measure the extent of the X-ray emission, we compared the surface brightness of the source  with what would be expected from a point source: if the surface brightness of the source exceeds that of a point source over a range of distances, then we can confirm that the source is extended and measure this spatial extent. We perform this exercise separately for the soft and hard band. In each band, we extracted radial surface brightness profiles from the soft and hard bands by calculating the background-subtracted net counts per pixel squared within a series of annuli around the nucleus. The radius of the inner annulus was set to 1$^{\prime\prime}$ and the subsequent annuli were incremented in steps of 0$\farcs$5.

\begin{figure}
  \centering
  \includegraphics[scale=0.28,angle=90]{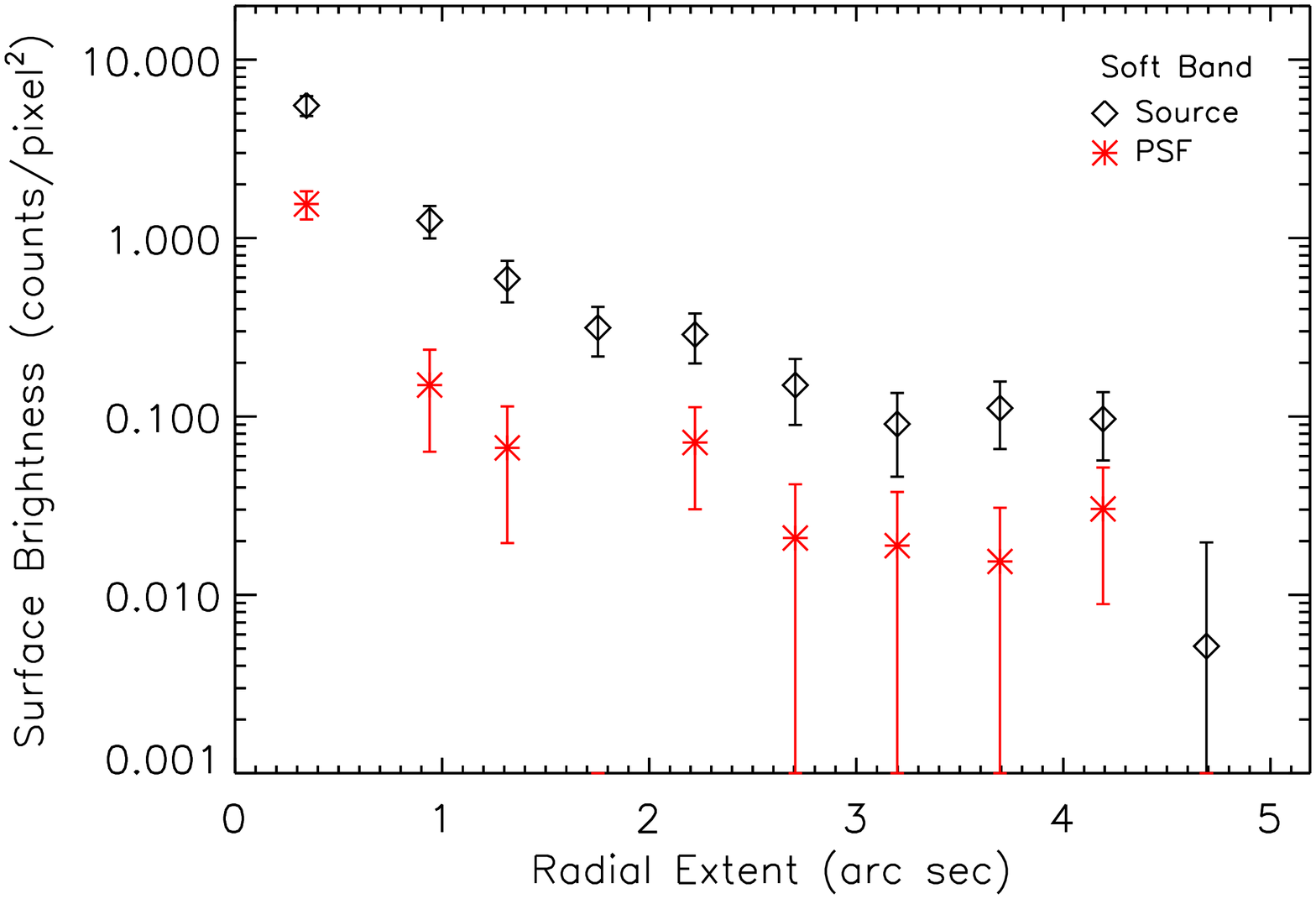}
  \includegraphics[scale=0.28,angle=90]{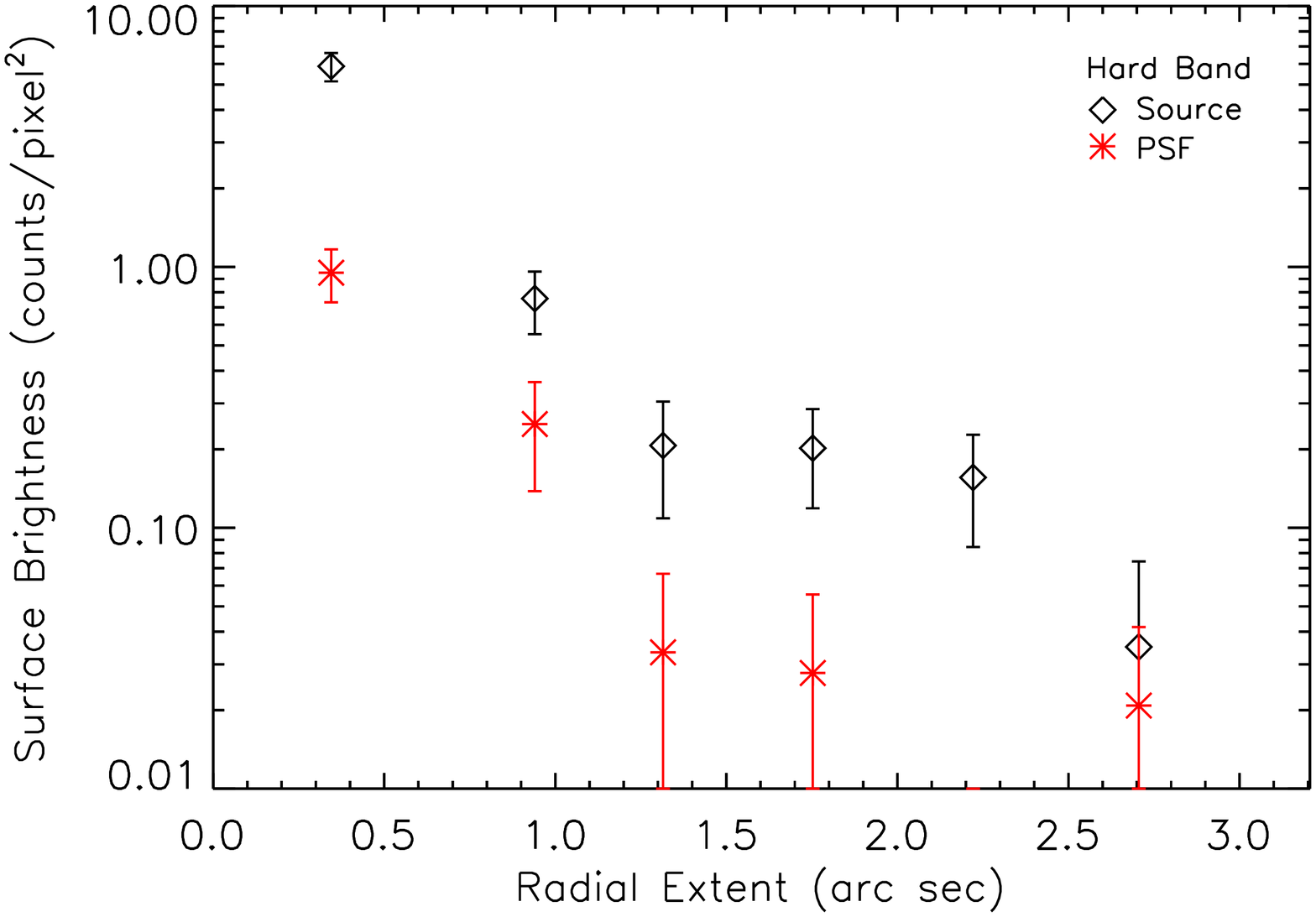}
  \caption{\label{extent} Radial surface brightness profiles of NGC 4968 (black diamonds) in the soft ({\it top}) and hard ({\it bottom}) bands compared with the radial surface brightness profiles for a simulated point source (red asterisks) at the source position. The source is significantly extended to $\sim$4$^{\prime\prime}$ (900 pc) in soft band and up to $\sim$2.25$^{\prime\prime}$ (500 pc) in the hard band.}
\end{figure}

In Figure \ref{extent}, we compare the radial surface brightness profiles from the data with that of the simulated point source. Figure \ref{extent} demonstrates that the soft X-ray emission is significantly extended up to $\sim4^{\prime\prime}$ while the hard X-ray emission is more compact, yet extends up to $\sim2.25^{\prime\prime}$. These sizes correspond to physical scales of 900 pc and 500 pc, respectively. As discussed in Section \ref{ext_emis}, the size scale, and direction of the extended hard emission is consistent with the narrow line region measured via the [OIII] ionization cone \citep{schmitt03}, while the soft emission appears to be independent of the narrow line region.

\section{Spectral Analysis}\label{spec}

The {\it Chandra} spectrum was fitted with \textsc{XSpec} version 12.9.0 \citep{arnaud}.  We used the Cash-statistic for fitting, which is more appropriate in the low count regime than $\chi^2$ since it only requires that the count distribution be Poissonian \citep{cash}. The energy range was restricted to be between 0.5 and 8 keV. Errors on the fit parameters represent the 90\% confidence interval for one interesting parameter.

\subsection{Global Fit}

\subsubsection{Phenomenological Fit}
\begin{figure}
  \centering
  \includegraphics[scale=0.35,angle=270]{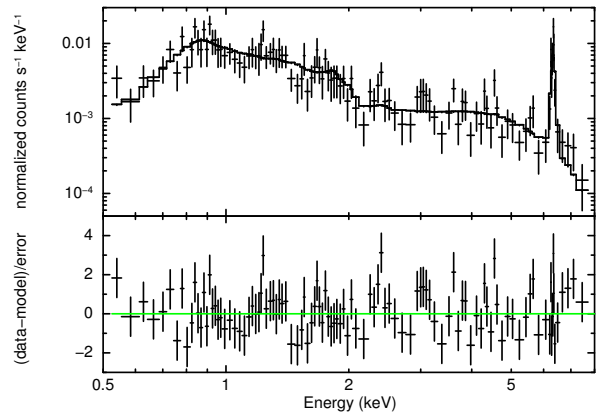}
  \caption{\label{empir_spec} {\it Chandra} spectrum fitted with a phenomenological model of a double absorbed power law with a Gaussian component to accommodate the Fe K$\alpha$ line and an \textsc{apec} component to best model emission at soft energies. From measuring the Fe K$\alpha$ EW against the total continuum, we find a value of 2.9 keV.}
  \end{figure}

We begin by fitting the {\it Chandra} spectrum with a phenomenological model to measure the equivalent width of the Fe K$\alpha$ feature as this provides insight into the geometry of the obscurer. A simple absorbed power law fails to fit the spectrum, requiring a more complex model. Here, we adopt a double absorbed power law model which describes a partial covering geometry where a portion of the intrinsic AGN continuum is scattered into our line of sight, or leaks through the circumnuclear obscuration, while the rest is absorbed. To this model, we add a Gaussian component, at the redshift of the source, to accomodate the Fe K$\alpha$ emission and a thermal emission model (\textsc{apec}) to best fit the soft emission (see below). One absorption component ($N_{\rm H}$) is frozen to the Galactic value of 9$\times10^{20}$ cm$^{-2}$ and an additional component is left free to attenuate the the transmitted AGN emission. This fit to the {\it Chandra} spectrum is shown in Figure \ref{empir_spec}, where we find $N_{\rm H}$=1.8$^{+5.0}_{-0.9}\times10^{23}$ cm$^{-2}$ and a power law slope ($\Gamma$) of 1.69$^{+0.47}_{-0.51}$.

\begin{figure*}
  \centering
  \includegraphics[scale=0.45]{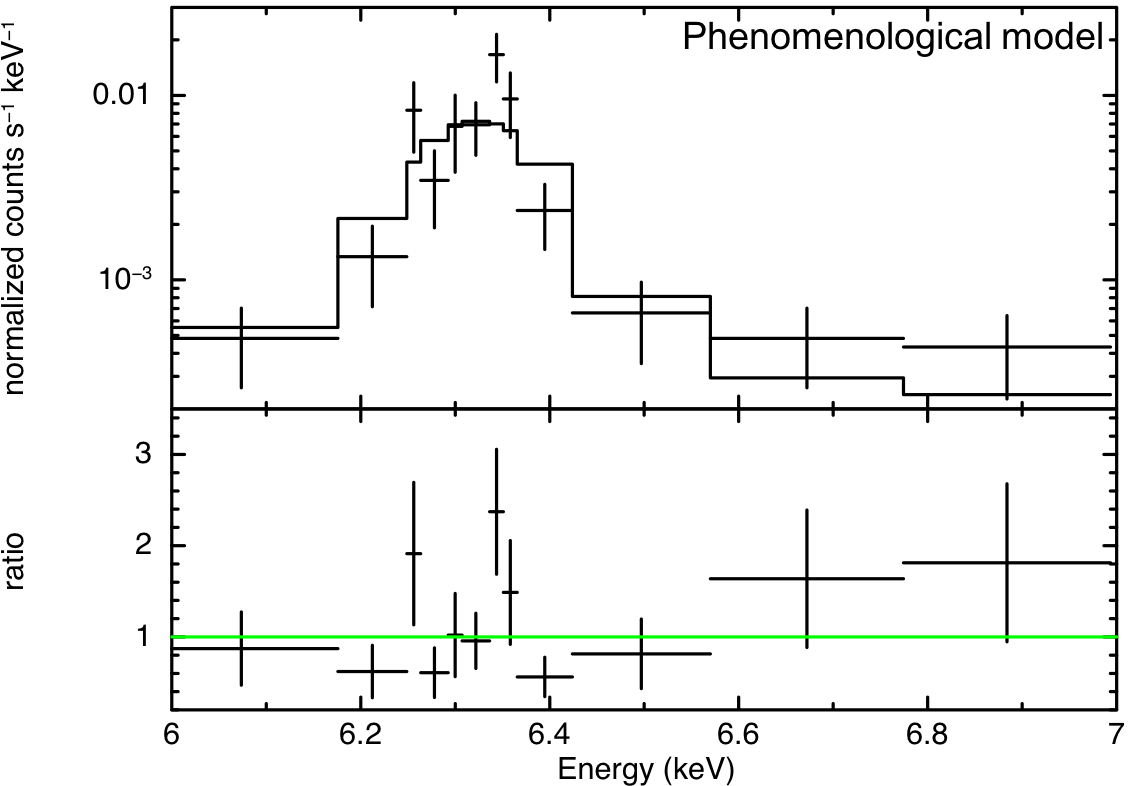}~
  \includegraphics[scale=0.45]{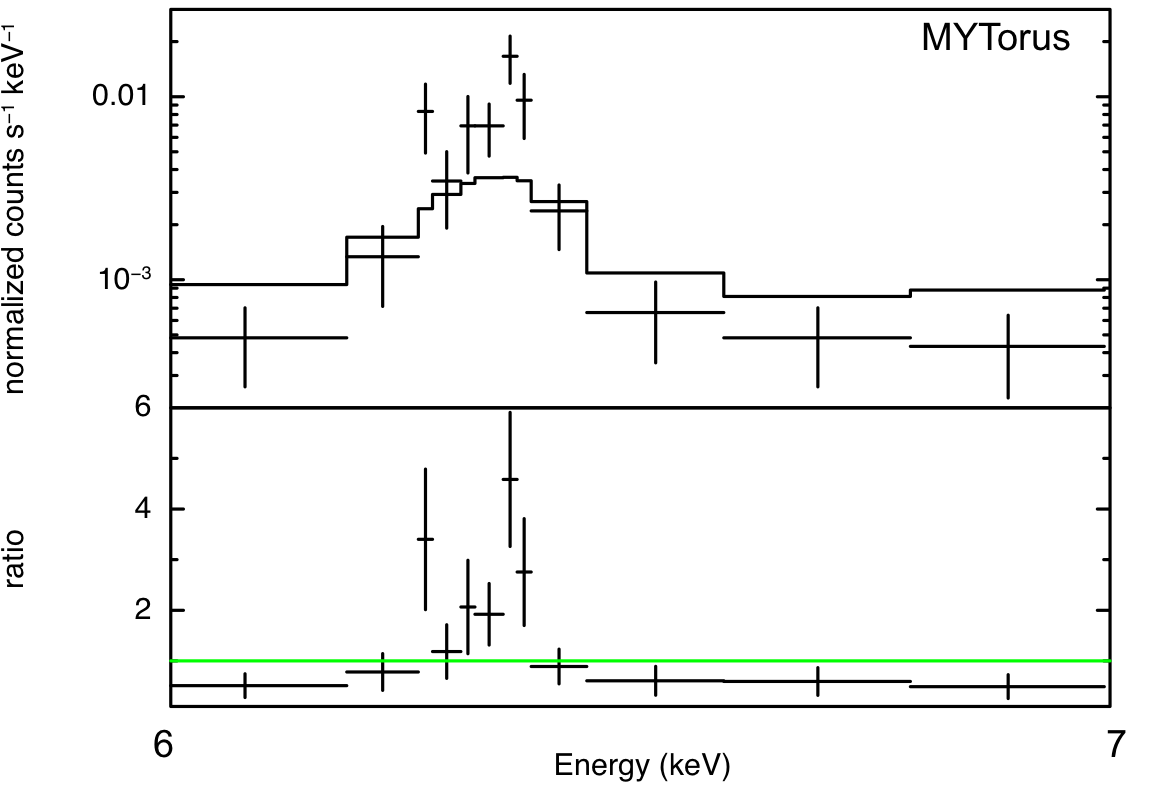}
  \includegraphics[scale=0.45]{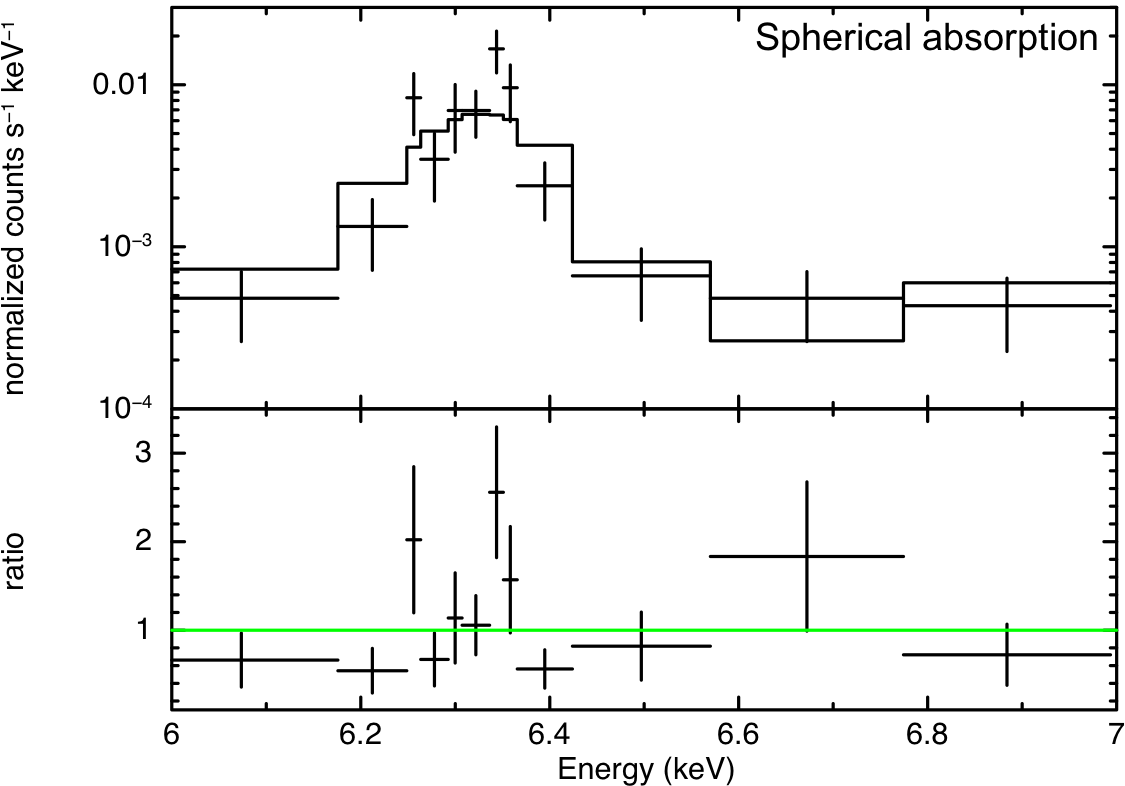}~
  \caption{\label{fe_ka} Close-up of the Fe K$\alpha$ complex for different model fits, with best-fit model in upper panels and fit residuals (i.e., the ratio of data to the model) in bottom panels. The phenomenological model ({\it top left})  fits the extreme Fe K$\alpha$ emission feature, though it is unable to accommodate all the data points; the line is unresolved (i.e., the best fit width of the line is pegged at the lower limit which is set to several times below the instrumental resolution).  The physically motivated MYTorus model ({\it top right}) does a poorer job of fitting the emission line than the phenomenological model and the \citet{brightman} spherical absorption model ({\it bottom}). The scenario where the AGN continuum is nearly completely cocooned within Compton-thick material appears the most likely description of the data.}
\end{figure*}

In Figure \ref{fe_ka} (top left), we show a close-up of the Fe K region and note that this model can not completely fit the line, despite the line being spectrally unresolved (i.e., the best-fit $\sigma$ pegged at our imposed lower limit of 0.01 keV, which is several times below the approximate {\it Chandra} resolution of $\sim$0.06 keV); this could be due to poor photon statistics in the line region, causing the residuals to be more narrow than the spectral line resolution. We measured the Fe K$\alpha$ EW against the total continum, finding a value of 2.9 keV in the rest-frame.

\begin{figure}
  \centering
  \includegraphics[scale=0.37]{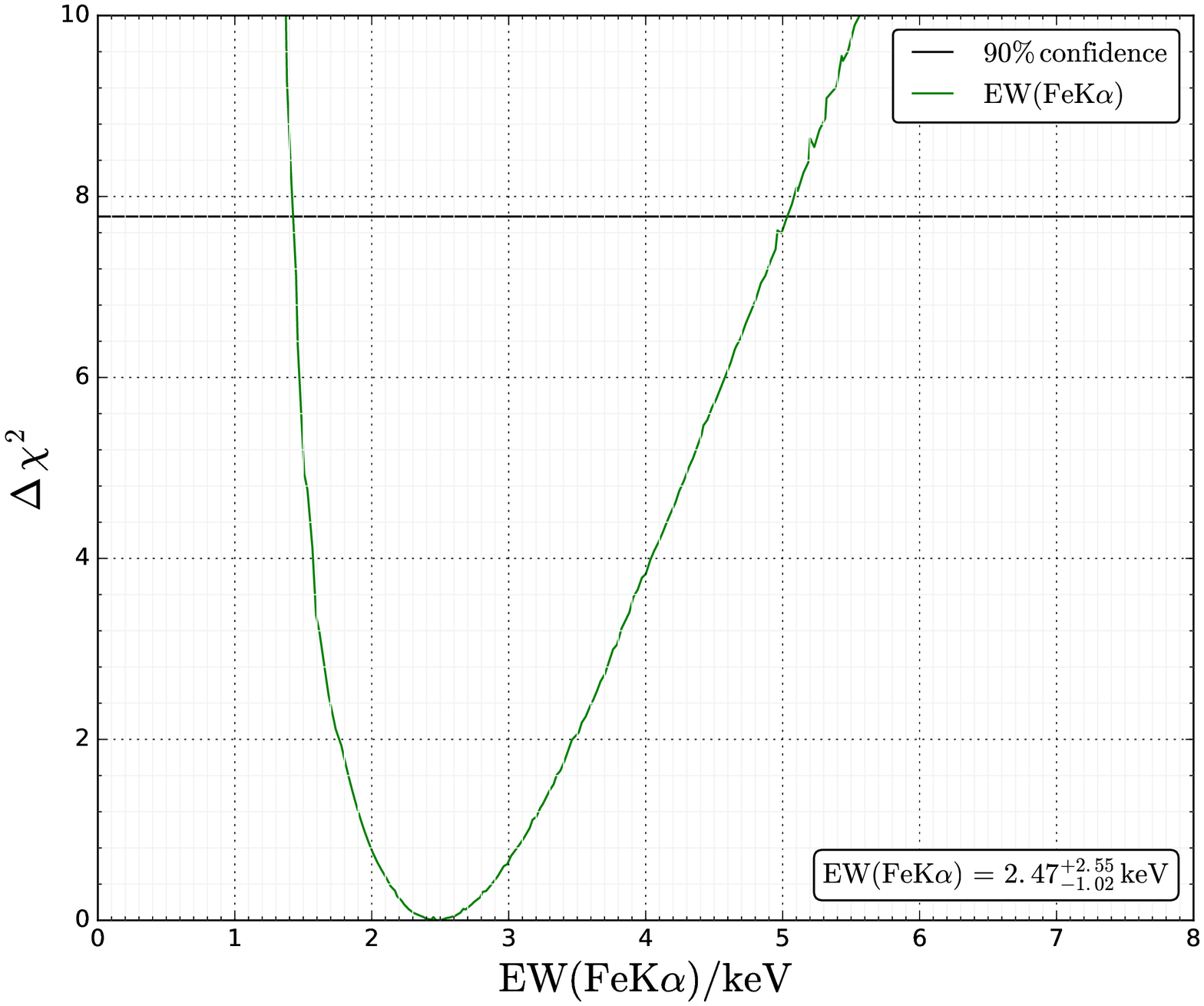}
  \caption{\label{loc_ew}  Fe K$\alpha$ EW as a function of $\Delta \chi^2$, calculated over a four dimensional grid of power law slope and normalization and Gaussian width and normalization. The horizontal black line indicates the 90\% confidence level for four free parameters ($\Delta \chi^2$ = 7.779). The large EW indicates that the inclination angle of the obscurer is completely edge-on, and that the opening angle of the torus may approach 0$^{\circ}$, such that the AGN is completely enshrouded.}
\end{figure}

We also perform a local fit (3.5 keV - 7.8 keV) for a more detailed investigation of the Fe K$\alpha$ EW. Here, the spectrum is modeled with a power law and Gaussian component. Taking into account errors on the local power law slope and normalization, and width and normalization of the Gaussian model (the line center was well measured at 6.38 keV and thus was frozen), we calculated $\Delta$ $\chi^2$ as a function of Fe K$\alpha$ EW in Figure \ref{loc_ew}; the horizontal line denotes the 90\% confidence level ($\Delta \chi^2$= 7.779 for four free parameters). From this exercise, we find the (rest-frame) Fe K$\alpha$ EW = 2.5$^{+2.6}_{-1.0}$ keV.

This large EW is quite telling. To obtain such large values, the transmitted emission must be attenuated more than the fluorescent line emission produced within the medium, which is possible when the column density reaches Compton-thick levels of obscuration. The same effect could be induced by variability, where the AGN continuum weakens and the travel time to the Fe K$\alpha$ emitting region causes a delay in the response to the EW. As we discuss further in Section \ref{xray_var}, archival X-ray observations of this source indicate the flux and EW have remained constant over time (20 and 15 years, respectively), such that this scenario seems unlikely. The measured EW thus implies a higher $N_{\rm H}$ than what is measured, indicating that another continuum component must dominate at energies below the Fe K$\alpha$ line. Additionally, as discussed in \citet{mytorus}, the Fe K$\alpha$ EW is sensitive to the inclination angle of the obscuring medium when the line-of-sight intersects the torus, with the most extreme EW values (i.e., $>$2 keV) obtained when the torus is completely edge-on at 90$^{\circ}$  for the geometry assumed in the MYTorus model, i.e., a torus with a fixed 60$^{\circ}$ opening angle.

\subsubsection{Toroidal Obscuration}
To obtain a physically motivated fit to the NGC 4968 spectrum, we use the \textsc{MYTorus} model which self-consistently treats the attenuated transmitted continuum (modeled as a powerlaw), reflection spectrum, and Fe K$\alpha$  and K$\beta$ fluorescence emission \citep{mytorus}. To preserve the self-consistency of the model, the normalization and $\Gamma$ of the powerlaw, the inclination angle of the torus, and the column density ($N_{\rm H}$) are tied together among the \textsc{MYTorus} model components. Motivated by our phenomenological modeling of the spectrum and resulting Fe K$\alpha$ EW measurement, we freeze the inclination angle to 90$^{\circ}$.

\begin{figure*}
  \centering
  \includegraphics[scale=0.45]{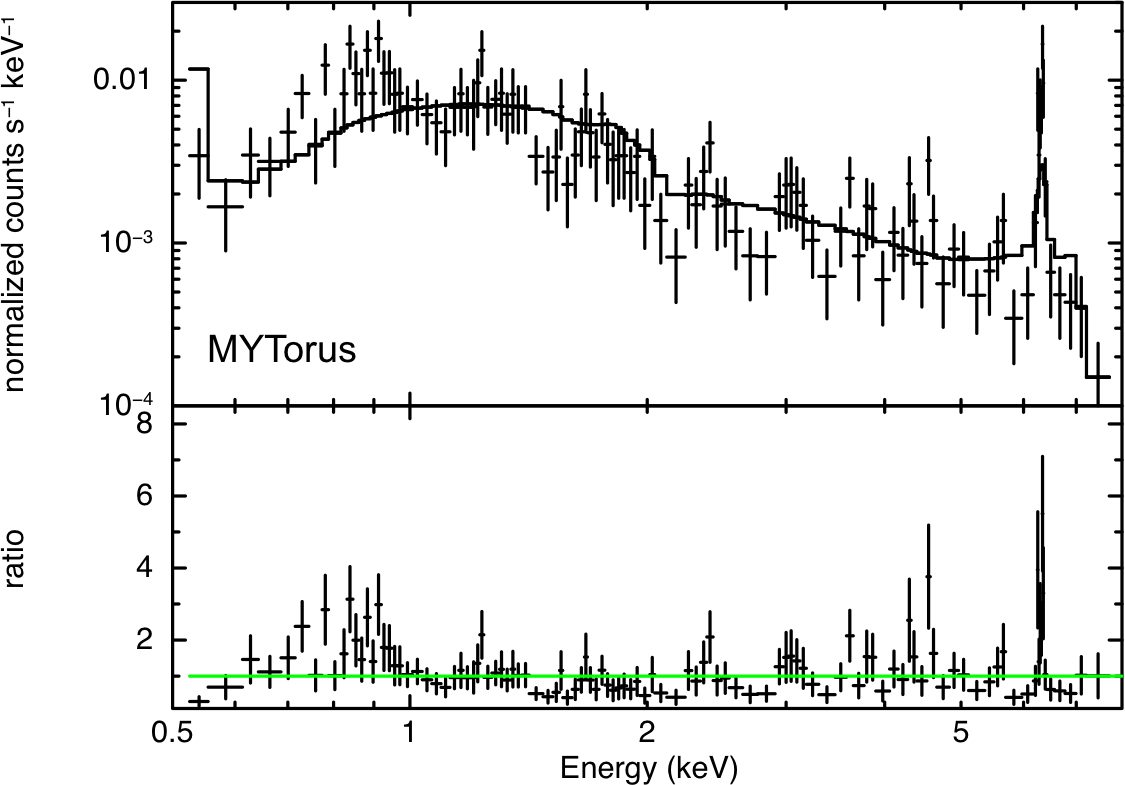}~
  \includegraphics[scale=0.45]{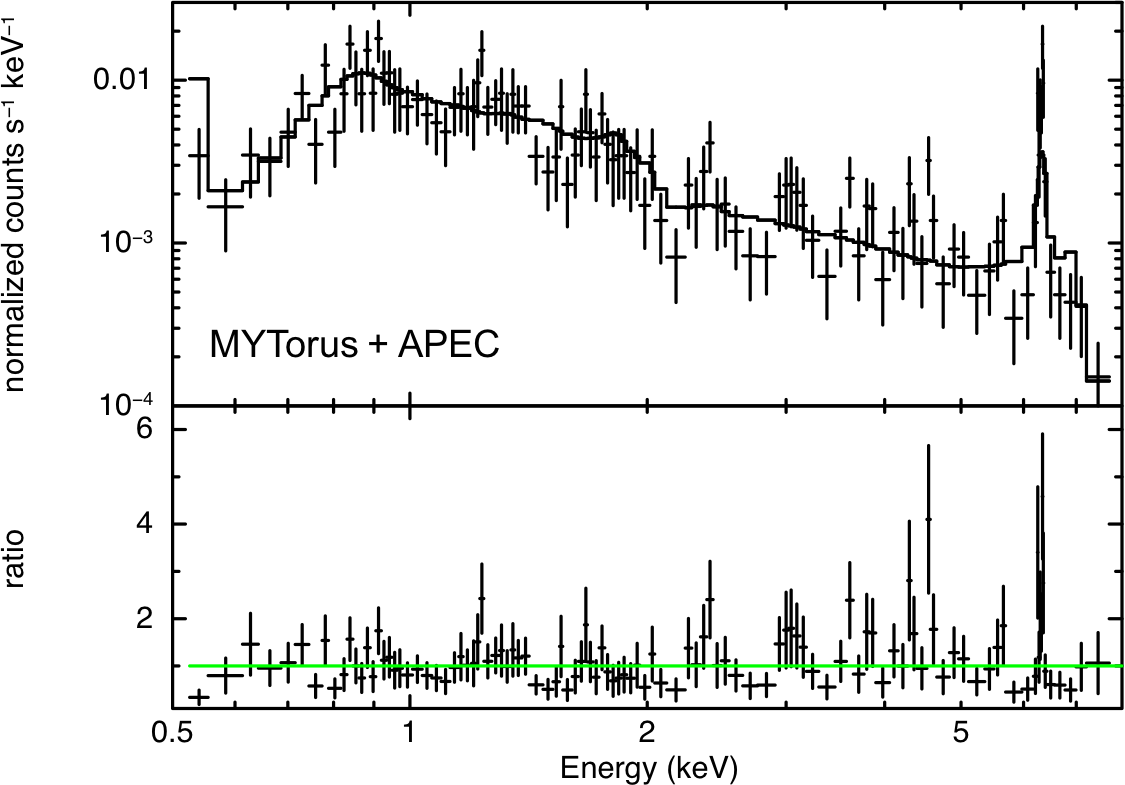}
  \includegraphics[scale=0.45]{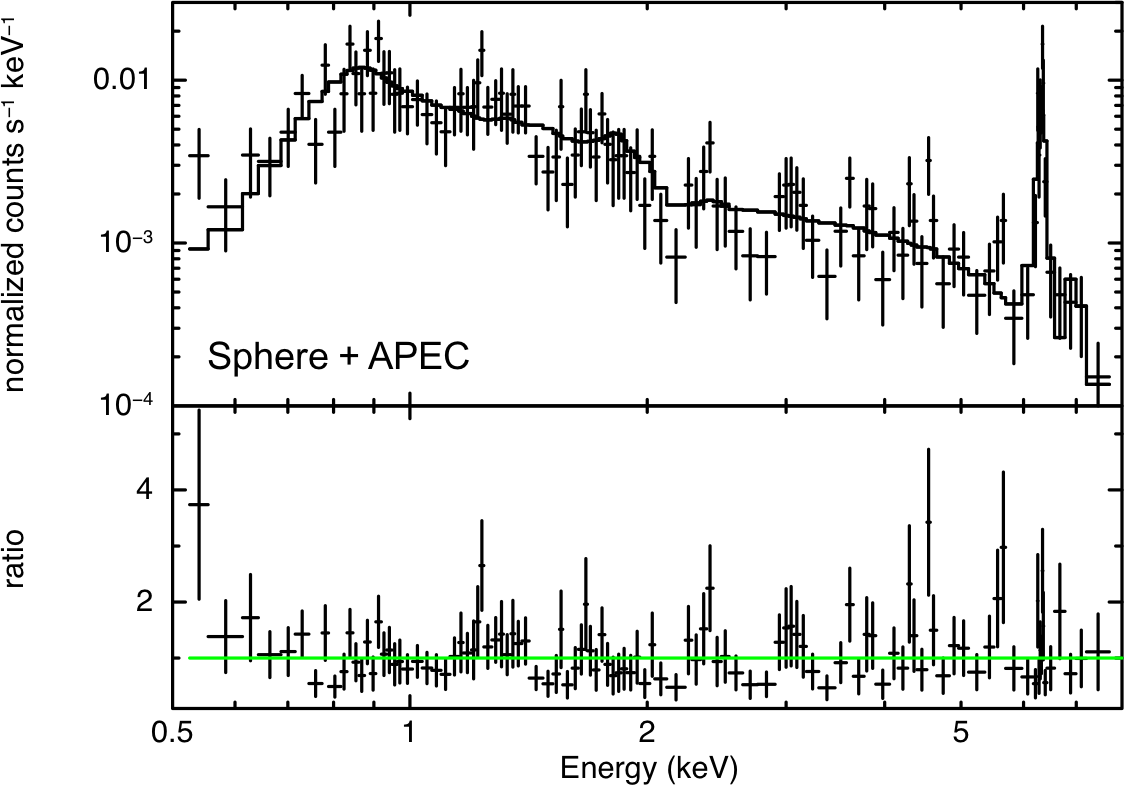}
  \caption{\label{fig_spec_global} {\it Chandra} spectrum (black) of the global X-ray emission fitted with \textsc{MYTorus} only ({\it top left}) and with \textsc{MYTorus} plus \textsc{apec} ({\it top right}). In the {\it bottom panel}, the spherical absorption model of \citet{brightman} is used, again with an \textsc{apec} component added to best accommodate emission at softer energies. All models include a component where a fraction of the AGN continuum emission is scattered into the observer line of sight. The \textsc{apec} component best fits the features at lower energies, indicating that a portion of the soft emission is thermal, putatively linked to circumnuclear star formation. }
\end{figure*}

The data require an additional powerlaw component to properly model the soft emission (i.e., $\Delta$C-Stat=703 for a change of 1 in degrees of freedom (DOF) between the model lacking this component and the one where it is included). We attribute this component to be due to a portion of the intrinsic AGN continuum that is scattered into or directly enters our line of sight after leaking through the torus: the powerlaw normalization and $\Gamma$ are tied to the \textsc{MYTorus} components, with a constant factor left free to account for the scattered/leaked fraction of AGN light. In Figure \ref{fig_spec_global} (top left), we show this fit to the {\it Chandra} spectrum, with the model parameters summarized in the first column of Table \ref{spec_global}. Residuals at softer energies are apparent.

We therefore added an \textsc{apec} component to this model which accounts for thermal emission, putatively associated with star formation; the abundances are kept frozen at solar. The lower energy spectrum is much better modeled ($\Delta$C-Stat = 60.3 for $\Delta$DOF = 2, corresponding to $P < 0.005$ that the statistical improvement to the fit is due to chance). However, the Fe K$\alpha$ complex is worse fit when compared with the phenomenological model, with larger fit residuals around the emission feature (Figure \ref{fe_ka}). We note that in this modeling, we used MYTorus tables which have a power law termination energy of 200 keV. However when we fit the spectra with termination energies of 100 and 400 keV, which span the range of the highest and lowest values available with \textsc{MYTorus}, we obtain essentially identical results.

\begin{deluxetable*}{llll}
  \tablewidth{0pt}
  \tablecaption{\label{spec_global}Spectral Fit to Global Emission}
  \tablehead{\colhead{Parameter} &  \colhead{\textsc{MYTorus}} & \colhead{\textsc{MYTorus + apec}} &\colhead{Sphere + \textsc{apec}}} 
  \startdata
  $\Gamma$                                 & 1.57$^{+0.16}_{-0.12}$ & $<$1.43 & $<$1.27 \\
  $N_{\rm H}$ ($10^{24}$ cm$^{-2})$           & 1.36$^{+0.31}_{-0.34}$ & 1.54$^{+0.23}_{-0.26}$ & 7.64$^{+8.57}_{-2.37}$ \\
  Power law normalization (10$^{-3}$)\tablenotemark{1} & 1.81$^{+0.79}_{-0.63}$ & 1.83$^{+0.83}_{-0.63}$ & 2.85$^{+0.23}_{-1.96}$\\
  $f_{\rm scatt}$ (10$^{-2}$)\tablenotemark{2}           & 1.3$^{+0.8}_{-0.4}$ & 0.9$^{+0.5}_{-0.3}$ & 0.45$^{+0.05}_{-0.34}$\\
  kT (keV)                                 &  \nodata                & 0.72$^{+0.12}_{-0.11}$ & 0.72$^{+0.10}_{-0.11}$ \\
  \textsc{apec} normalization (10$^{-6}$)\tablenotemark{3} & \nodata   & 6.77$^{+1.68}_{-1.59}$ & 8.71$^{+1.95}_{-1.42}$\\
  C-Stat (DOF)                             & 217.2 (106) & 156.9 (104) & 128.5 (104)
  \enddata
  \tablenotetext{1}{Power law normalization is units of photons keV$^{-1}$ cm$^{-2}$ s$^{-1}$ at 1 keV.}
  \tablenotetext{2}{Fraction of intrinsic AGN light that leaks through the circumnuclear obscuration or is scattered into the observer line of sight.}
  \tablenotetext{3}{The \textsc{apec} normalization is given as $\frac{10^{-14}}{4\pi[D_A(1+z)]^2}$ $\int n_e n_h dV$, where $D_A$ is the angular diameter distance to the source in cm and $n_e$ and $n_H$ are the electron and Hydrogen densities, respectively, in cm$^{-3}$.}
  \end{deluxetable*}

\subsubsection{Spherical absorption}
To produce such an extreme Fe K$\alpha$ EW, the central engine may be completely enshrouded, because differential extinction of continuum and fluorescent line photons is greater for a closed geometry such as a uniform spherical distribution, compared to an open geometry such as a torus (e.g., compare left panel of Figure 2 from \citealp{brightman} with Figure 8 from \citealp{mytorus}). We therefore fit the spectra with the spherical absorption model of \citet{brightman}.

Similar to the MYTorus model, the spherical absorption model treats the transmitted, reflected and fluorescent line emission self-consistently. In this model, the central continuum is fully covered over 4$\pi$ steradian, though we do include a scattered powerlaw component to account for light leaking through this obscuration, similar to the toroidal model; the percentage of escaping light is small enough to not significantly affect the self-consistency of the model. Again, we include an \textsc{apec} component to best fit the soft emission ($\Delta$C-Stat = 48.2 for $\Delta$DOF = 2, corresponding to $P < 0.005$). The bottom panel of Figure \ref{fig_spec_global} shows the spherical absorption fit to the X-ray spectra of NGC 4968, with a close-up of the Fe K complex shown in Figure \ref{fe_ka}. This model provides a better fit to the data than the toroidal geometry; we note that thawing the Fe abundance does not improve the fit and leads to an abundance that is 10 times solar, which is unphysical. The fit parameters for the spherical absorption model are listed in Table \ref{spec_global}.

\subsubsection{Summary of the {\it Chandra} Spectral Modeling to the Global Emission}
The spectral modeling points to the AGN in NGC 4968 being heavily obscured to Compton-thick. The implied column density of the obscurer is inconsistent between the \textsc{MYTorus} and spherical absorption models, ranging from N$_{\rm H} = 1.54^{+0.23}_{-0.26} \times 10^{24}$ cm$^{-2}$ (marginally Compton-thick) to $7.64^{+8.57}_{-2.37} \times 10^{24}$ cm$^{-2}$ (extending to extreme Compton-thick levels). Furthermore, circumnuclear star formation is on-going in the host galaxy, evidenced by the thermal nature of the soft X-ray spectrum.

We note that both model fits return a shallow spectral slope, inconsistent with the canonical $\Gamma$=1.8-1.9 for AGN \citep[e.g.,][]{kim,tozzi}. However, if we freeze the photon index to 1.9, we obtain statistically worse fits ($\Delta$C-Stat = 53.5, 40.7 for $\Delta$DOF = 1, for the MYTorus and spherical absorption models, respectively), with larger residuals around the Fe K complex than when the photon index is a free parameter. Furthermore, a hard spectrum facilitates the production of extreme Fe K$\alpha$ EW values: with more photons in the continuum above the Fe K edge, more Fe K$\alpha$ photons can be produced \citep[see][]{mytorus}.

From these fits, we calculated the total observed soft and hard X-ray luminosities ($L_{\rm 0.5-2 keV}$ and $L_{\rm 2-10keV}$, respectively), the \textsc{apec} contribution to the soft X-ray luminosity ($L_{\rm thermal}$) from which we estimate the star formation rate in Section \ref{sfr}, and the estimated instrinsic 2-10 keV X-ray luminosity ($L_{\rm 2-10keV, int}$) from the power law fit parameters. These luminosities, and associated errors (which reflect the statistical errors on the powerlaw and/or \textsc{apec} normalizations) are listed in Table \ref{lum}. Both models return similar observed X-ray luminosities and predict intrinsic 2-10 keV luminosities that are over two orders of magnitude larger than that observed.

\begin{deluxetable}{lll}[l]
  \tablewidth{0pt}
  \tablecaption{\label{lum}X-ray Luminosities\tablenotemark{1}}
  \tablehead{\colhead{Luminosity} & \colhead{MYTorus} & \colhead{Sphere}}
  \startdata
  $L_{\rm 0.5-2 keV}$ (observed) & 40.11$^{+0.19}_{-0.23}$ & 40.10$^{+0.10}_{-0.53}$ \\
  $L_{\rm 2-10keV}$ (observed)   & 40.91$^{+0.17}_{-0.18}$ & 40.77$^{+0.04}_{-0.51}$ \\
  $L_{\rm thermal}$              & 39.67$^{+0.10}_{-0.12}$ & 39.78$^{+0.09}_{-0.08}$ \\
  $L_{\rm 2-10keV, int}$          & 42.45$^{+0.17}_{-0.18}$ & 42.89$^{+0.04}_{-0.51}$ 
  \enddata
  \tablenotetext{1}{Luminosities are in log space and units of erg s$^{-1}$ and are rest-frame values. The quoted errors on the luminosities refer to the statistical error of the fit.}
\end{deluxetable}
\vspace{1cm}

\subsubsection{Clues From the Mid-Infrared}
The scenario where the source is completely obscured geometrically, with a small amount of leakage permitted, appears to be the most plausible for explaining the X-ray spectrum of NGC 4968: the residuals around the Fe K complex are the lowest among the self-consistent models, as would be expected if complete coverage facilitates conditions to produce an extreme Fe K$\alpha$ EW. This model also predicts the highest column density.

Further clues of extreme obscuration are evidenced by the mid-infrared (MIR) emission of this source. The 12$\mu$m luminosity of NGC 4968 as measured by {\it WISE} \citep{wright} is 1.7$\times10^{43}$ erg s$^{-1}$. Based on the correlation found between the nuclear 12$\mu$m luminosity and intrinsic 2-10 keV X-ray luminosity \citep[log($\frac{L_{\rm 12\mu m}}{{\rm 10^{43} erg s^{-1}}}{\rm )} = (0.33 \pm 0.04) + (0.97 \pm 0.03) \rm{log(}\frac{L_{\rm 2-10keV, int}}{10^{43} erg s^{-1}}{\rm )}$;][]{asmus,gandhi}, the predicted X-ray luminosity is $\sim8 \times10^{42}$ erg s$^{-1}$, fully consistent with the value we derived via fitting the X-ray spectrum with the spherical absorption model. Though the MIR - X-ray correlation can also be driven by the X-ray spectral slope and covering factor of the obscurer \citep{yaqoob2011}, the MIR $W1 - W2$ color is 0.89 (Vega),\footnote{ We use the {\it WISE} magnitudes measured with profile-fitted photometry.} consistent with the MIR emission being dominated by AGN heated dust \citep[$W1-W2 > 0.8$ for $W2\leq 15$;][]{stern}. A high quality X-ray spectrum above 10 keV from e.g., {\it NuSTAR}, would determine whether a spherical or toroidal geometry is correct, since the predicted spectral shapes above 10 keV differ between these models.

\subsection{Nuclear Region}
Here we capitalize on {\it Chandra}'s arcsecond resolution to hone in on nuclear emission at scales of $\sim$1.7$^{\prime\prime}$ ($<$400 pc). As discussed above, a model where the AGN continuum is completely enshrouded is the most likely description of this source. In addition to fitting the nuclear spectrum with the spherical absorption model, we also applied \textsc{MYTorus} and summarize those results in Table \ref{spec_nuclear} for completeness.

\begin{figure}
  \centering
  \includegraphics[scale=0.45]{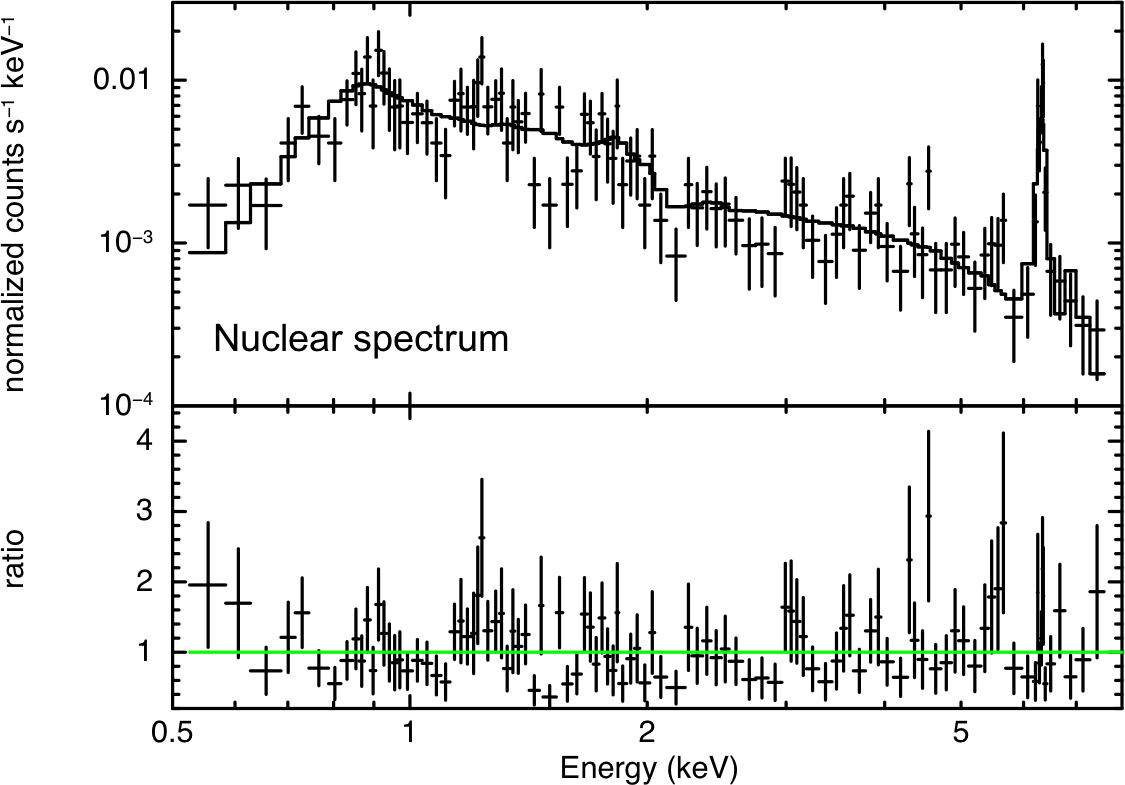}
  \caption{\label{fig_spec_nuc} {\it Chandra} spectrum of nuclear emission, fit with the \citet{brightman} spherical absorption model with a scattered AGN component (parameterized by a powerlaw model) and \textsc{apec} component. Thermal emission appears to be present on nuclear, as well as global, scales (see Figure \ref{fig_spec_ext}).}
\end{figure}

For both models, where we include a scattered powerlaw model to accommodate leaked AGN emission, we again find residuals at softer energies, such that adding in an \textsc{apec} component significantly improves the fit (the change in C-Stat is significant at the $P < 0.005$ level; Table \ref{spec_nuclear}). Thus it appears that thermal X-ray emission is not confined to the extended emission, but also arises at nuclear scales. The spherical absorption plus \textsc{apec} fit is shown in Figure \ref{fig_spec_nuc}.

\begin{deluxetable*}{lllll}[l]
  \tablewidth{0pt}
  \tablecaption{\label{spec_nuclear}Spectral Fit to Nuclear Emission}
  \tablehead{\colhead{Parameter} &  \colhead{\textsc{MYTorus}} & \colhead{\textsc{MYTorus + apec}} &\colhead{Sphere} &\colhead{Sphere + \textsc{apec}} }
  \startdata

  $\Gamma$                           & $<$1.59 & $<$1.44 & 1.55 $^{+0.14}_{-0.14}$ & $<$1.18 \\
  $N_{\rm H}$ ($10^{24}$ cm$^{-2})$       & 1.36$^{+0.30}_{-0.33}$ & 1.36$^{+0.26}_{-0.26}$ & 2.00$^{+1.12}_{-1.00}$ & 2.8$^{+13.6}_{-1.2}$ \\
  Power law normalization (10$^{-3}$)           & 1.47$^{+0.74}_{-0.54}$ & 1.42$^{+0.72}_{-0.55}$ & 3.62$^{+6.38}_{-2.65}$ & 1.25$^{+2.15}_{-0.75}$ \\
  $f_{\rm scatt}$ (10$^{-2}$)      & 1.4$^{+0.9}_{-0.5}$ & 1.2$^{+0.8}_{-0.4}$ & 0.64$^{+1.8}_{-0.4}$ & 1.0$^{+1.5}_{-0.6}$\\
  kT (keV)                           & \nodata & 0.76$^{+0.14}_{-0.14}$ & \nodata & 0.76$^{+0.12}_{-0.12}$ \\               
  \textsc{apec} normalization (10$^{-6}$)  & \nodata & 4.63$^{+1.58}_{-1.46}$ & \nodata & 6.52$^{+1.55}_{-1.74}$\\
  C-Stat (DOF)                       & 173.3 (101) & 141.0 (99) & 146.9 (101) & 112.7 (99) 

  \enddata
  \end{deluxetable*}

\subsection{Extended Emission}
\begin{figure}
  \centering
  \includegraphics[scale=0.45]{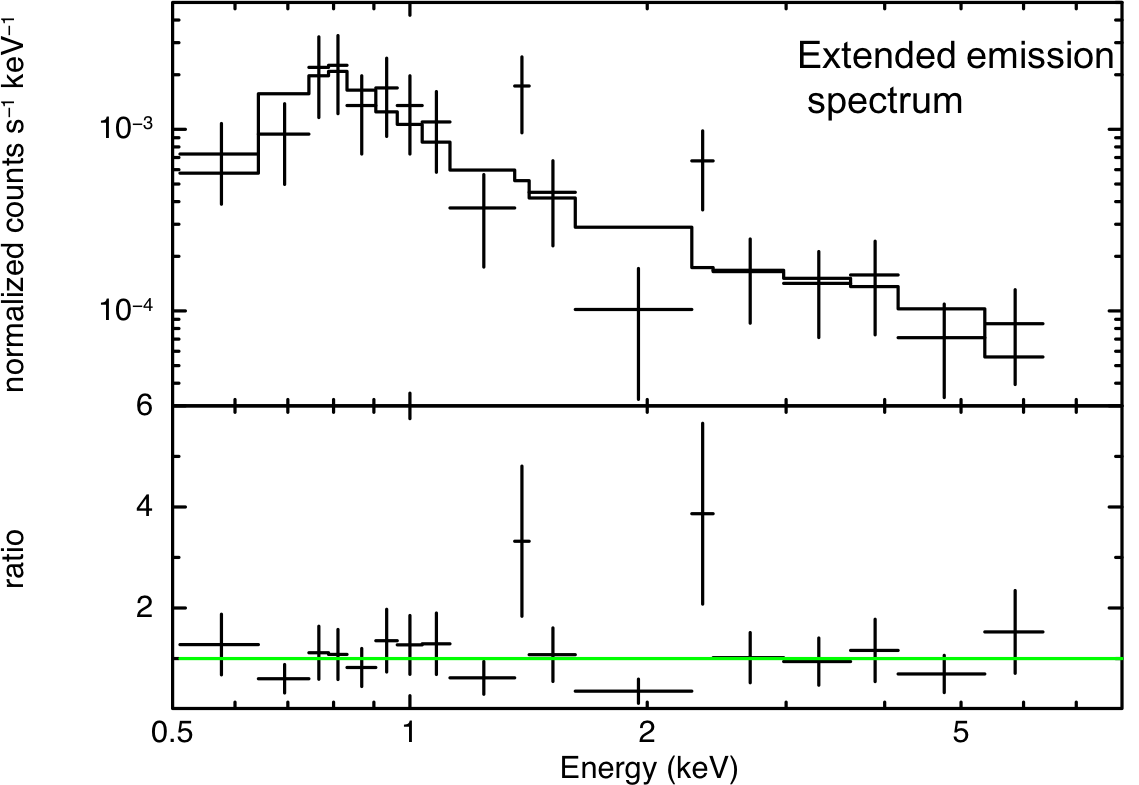}
  \caption{\label{fig_spec_ext} {\it Chandra} spectrum of extended emission, fit with an \textsc{apec} plus powerlaw model. The spectrum is soft, consistent with being dominated by thermal emission likely arising from star formation.}
\end{figure}

The spectrum from the extended only emission, extracted from a 6$^{\prime\prime}$ ($\sim$1.3 kpc) region with the central point source excised, is shown in Figure \ref{fig_spec_ext}. As expected from the imaging analysis, the spectrum is dominated by emission at soft energies. It is well accomodated by an \textsc{apec} plus powerlaw model (see Table \ref{spec_ext}), where both components are required to fit the spectrum. The powerlaw emission can arise from unresolved X-ray binaries in the host galaxy and/or AGN continuum photons that scatter into our line of sight.

In this extended zone, the temperature of the gas is significantly lower than that observed on nuclear scales with kT = 0.33$^{+0.19}_{-0.08}$ keV versus 0.76$^{+0.12}_{-0.12}$ keV. In fact, from the global fit to the X-ray emission, we find a gas temperature (kT $\sim$ 0.7) consistent with the nuclear value. Additionaly, the thermal luminosity is higher on nuclear than extended scales (4.33$^{+1.03}_{-1.15} \times 10^{39}$ erg s$^{-1}$ and 1.77$^{+1.04}_{-0.80} \times 10^{39}$ erg s$^{-1}$, respectively), suggesting that circumnuclear star formation dominates the total thermal emission in this galaxy. 

\begin{deluxetable}{ll}
  \tablewidth{0pt}
  \tablecaption{\label{spec_ext}Spectral Fit to Extended Emission}
  \tablehead{\colhead{Parameter} &  \colhead{\textsc{apec + powerlaw}} }
  \startdata
  kT (keV) & 0.33$^{+0.19}_{-0.08}$ \\
  \textsc{apec} normalization (10$^{-6}$) & 4.40$^{+2.58}_{-2.00}$ \\ 
  $\Gamma$ &  0.69$^{+0.69}_{-0.75}$ \\
  Power law normalization (10$^{-7}$) & 9.39$^{+9.62}_{-5.70}$ \\
  C-Stat (DOF) & 19.7 (14)
  \enddata
  \end{deluxetable}

\section{Discussion}\label{disc}

\subsection{X-ray History of NGC 4968}\label{xray_var}
NGC 4968 was detected by the {\it HEAO} A1 survey \citep{polletta}, though was undetected by {\it GINGA} and the {\it ROSAT} All-Sky Survey \citep{bianchi}. \citet{turner} found NGC 4968 to be about five times dimmer in an {\it ASCA} observation compared with the previous {\it HEAO} A1 detection. It was thus targeted as a potential X-ray ``changing-look'' AGN with {\it XMM-Newton}, where it was thought that perhaps a change in column density drove the flux variation. However, as \citet{bianchi} report, two epochs of {\it XMM-Newton} observations from 2001 and 2004 revealed no variability. Indeed, the observed 2-10 keV flux with {\it XMM-Newton} ($\sim$3$\times 10^{-13}$ erg cm$^{-2}$ s$^{-1}$) is consistent with the reported {\it ASCA} flux ($\sim$4$\times 10^{-13}$ erg cm$^{-2}$ s$^{-1}$) from a 1994 observation. \citet{bianchi} concluded that the {\it HEAO} A1 flux, extrapolated from the observed count rate between 1-20 keV due to loss of spectral resolution resulting from a hardware failure, was likely erroneous.

From our {\it Chandra} spectrum, we find a 2-10 keV flux (3.5$^{+3.3}_{-1.2} \times 10^{-13}$ erg cm$^{-2}$ s$^{-1}$) consistent with our previously reported {\it XMM-Newton} flux \citep[2.1$\pm 0.3 \times 10^{-13}$ erg cm$^{-2}$ s$^{-1}$;][]{me2011}, as well as similar Fe K$\alpha$ EWs ( 2.5$^{+2.6}_{-1.0}$ keV and 3.1$^{+1.0}_{-0.8}$). \citet{fukazawa} report an Fe K$\alpha$ EW of 1.8$\pm$0.6 keV from fitting the {\it Suzaku} spectrum obtained in 2006, which is consistent with our past {\it XMM-Newton} and current {\it Chandra} measurements.\footnote{NGC 4968 was not detected at energies above 10 keV with {\it Suzaku}, precluding us from using higher energy spectra to constrain our spectral fitting.} Hence it appears that the X-ray flux and spectrum of NGC 4968 has remained relatively constant throughout at least two decades of observations, indicating that the extreme Fe K$\alpha$ EW is due to obscuration rather than time lags between the continuum and the Fe K$\alpha$ emitting region.

\subsection{Comparison of Torus Properties with Past X-ray and Infrared Studies}
\citet{brightman} fitted the {\it XMM-Newton} spectrum of NGC 4968 with their spherical absorption model, obtaining a column density of log ($N_{\rm H}$ [cm$^{-2}$]) = 24.5$^{+0.9}_{-0.3}$ dex. We obtain similar column densities when fitting the {\it Chandra} spectrum from both the global and nuclear emission regions with the same model. However, we demonstrate that a scattered powerlaw is required in order to adequately model the data.

\citet{lira} fitted the MIR spectrum with a clumpy torus model \citep{nenkovaa,nenkovab}, finding the viewing angle of the torus to be between 75$^{\circ}$-90$^{\circ}$ at the 67\% confidence level. Consistent with our results from the X-ray modeling, the line-of-sight intersects the obscurer edge-on. They determine a dust line-of-sight extinction of $\sim$100 mag, which with their assumed gas-to-dust ratio of $N_{\rm H}/A_{V} = 1.79\times 10^{21}$ cm$^{-2}$ mag$^{-1}$, corresponds to $N_{\rm H} \sim 2 \times 10^{23}$ cm$^{-2}$. The finding of a lower column density from MIR observations is not surprising. First, this technique probes only dust, and dust-free gas within the sublimation radius may dominate the X-ray measured column density. Secondly, the MIR emission is due to the ensemble emission of all the central clouds and is not sensitive to the conditions along this specific line of sight.

Relatedly, it is entirely consistent that the 9.7$\mu$m silicate strength\footnote{$S_{\rm sil}$ = ln($f_{\rm obs}$(9.7$\mu$m)/$f_{\rm cont}$(9.7$\mu$m)), where $f_{\rm obs}$ is the flux density of the silicate feature and $f_{\rm cont}$ is the flux density of the local mid-infrared continuum \citep{spoon}.} is relatively weak, despite the large column density. \citet{wu} measured $S_{\rm sil}$ = -0.21 from the {\it Spitzer} spectrum, where $S_{\rm sil} <$ -0.5 in the cases of strong absorption. Silicate strength does not directly measure line-of-sight optical depth but is instead sensitive to the geometric and temperature distribution of the emitting and absorbing material \citep{levenson2007}. Indeed, only about 45\% of the Compton-thick AGN studied by \citet{goulding} showed strong silicate absorption, which they attribute to dust in the host galxay, not immediately associated with AGN. NGC 4968 provides further evidence that MIR-only diagnostics do not always reveal the extreme nature of Compton-thick AGN.

\subsection{Extended X-ray Emission is Not Coincident with Features Discovered at Other Wavelengths}\label{ext_emis}
NGC 4968 was observed with the {\it Hubble Space Telescope} WFPC2 camera to image the [OIII] emission which traces the narrow line region \citep{schmitt03}. The [OIII] emission, and by proxy the narrow line region, was found to extend 2.2$^{\prime\prime}$ along  position angle 40$^{\circ}$, with an extent of 1.3$^{\prime\prime}$ in the perpendicular direction. Additionally, high-resolution Very Large Array (VLA) observations at 8.46 GHz revealed a radio structure at a position angle of 85$^{\prime\prime}$, extending 2.3$^{\prime\prime}$. The locations of these regions are noted in the top panel of Figure \ref{image}.

While the narrow line region seems coincident with the slightly extended X-ray emission in the hard and soft bands along the northeast-southwest direction, the expanse of soft X-ray emission East of the nucleus is not coincident with the narrow line region nor the radio structure. Additionally, this soft X-ray emission is about twice the size of both the narrow line region and radio emission ($\sim$4$^{\prime\prime}$ compared with 2.2-2.3$^{\prime\prime}$). The size and location of this expanse, plus its apparent thermal nature, lends credence to the interpretation that it is associated with star formation rather than AGN activity that permeates beyond nuclear scales.

\subsection{The X-ray View of Star Formation}\label{sfr}
We use the thermal emission from the \textsc{apec} component as a starting point to estimate the star formation rate. As pointed out in \citet{me2012}, using solely the \textsc{apec} luminosity neglects the contribution from X-ray binaries which emit as powerlaws. However, the unresolved X-ray binary population can not be easily disentangled from the AGN powerlaw emission. In \citet{me2012}, we presented a template to adjust $L_{\rm thermal}$ to recover X-ray binary emission and more accurately estimate the star formation rate (SFR). Using a sample of 22 star forming galaxies from the XSINGS sample \citep[a {\it Chandra} survey of the SINGS sample,][]{sings}, we calculated the mean ratio of the X-ray emission from non-nuclear point sources to the total 0.5-2 keV emission ($R$). The 0.5-2 keV luminosity attributable to star-formation ($L_{\rm 0.5-2keV,SF}$) is then:
\begin{equation}
  L_{\rm 0.5-2keV,SF} = \frac{L_{\rm thermal}}{1-R},
\end{equation}
where $R = 0.51\pm0.26$.

Using the above expression and $L_{\rm thermal}$ = 6.0$\times10^{39}$ erg/s (Table \ref{lum}), we estimate that the X-ray luminosity associated with star-formation is 1.2$\pm 0.6 \times10^{40}$ erg s$^{-1}$ (taking into account errors on the fitted \textsc{apec} luminosity and $R$ value). Using the X-ray luminosity to SFR conversion presented in \citet{pereira-santaella}, where they combine ultraviolet and infrared data to account for both the unobscured and obscured star formation in a local sample of luminous infrared galaxies:
\begin{equation}
  SFR \ ({\rm M_{\sun} \  yr^{-1}}) = 3.4 \times 10^{-40} L_{\rm 0.5-2keV,SF}\ ({\rm erg\ s^{-1}}),
\end{equation}
we find a SFR of  4.1$\pm$2.0 M$_{\sun}$ yr$^{-1}$. We note that when using the \citet{ranalli} $L_{\rm 0.5-2keV}$/SFR relation, which is calibrated on infrared data only and thus neglects unobscured star formation and also assumes a different initial mass function than \citet{pereira-santaella}, we find a SFR of 2.6$\pm$1.3 M$_{\sun}$ yr$^{-1}$.

Additionally, we estimate the far-infrared SFR from the {\it IRAS} flux densities at 60$\mu$m and 100$\mu$m \citep[2.35 Jy and 3.75 Jy, respectively;][]{12um}. Using the far-infrared derived SFR calibration from \citet{kennicutt}:
\begin{equation}
  SFR\ ({\rm M_{\sun}\ yr^{-1}}) = 4.5 \times 10^{-44} L_{\rm FIR}\ ({\rm erg\ s^{-1}}),
\end{equation}
we obtain a SFR of 2.4 M$_{\sun}$ yr,$^{-1}$ which is essentially identical to our X-ray derived SFR using the \citet{ranalli} relation; both works assume the same initial mass function. We note that the \citet{kennicutt} SFR calibration includes mid-infrared wavelengths which we did not include here to minimize contamination from AGN-heated dust to our star formation estimates.

The mid-infrared spectroscopy provides an independent probe of the AGN/star-formation connection in NGC 4968. In \citet{me2010}, we reported the polycyclic aromatic hydrocarbons (PAHs) EW measurements at 11.3$\mu$m and 17$\mu$m for the 12$\mu$m Sy2 sample using {\it Spitzer} data. PAHs are associated with star formation though the presence of AGN heated dust dilutes the strength of the PAH EW: the stronger the AGN contribution to the MIR continuum, the weaker the PAH EW. We found the PAH EWs to be weak. We can infer that circumnuclear star formation is on-going by the detection of the PAH features and presence of thermal X-ray emission, though AGN heated dust dominates the mid-infrared emission, evidenced by both the MIR spectroscopy and {\it WISE} photometry.

\subsection{Extreme Fe K$\alpha$ EW: An Indication of More than Just Obscuration?}
In addition to providing insight into the geometry of the X-ray reprocessor, an extreme Fe K$\alpha$ EW ($>2$ keV) may be a signature of circumnuclear star formation. \citet{levenson} pointed out in their analysis of Compton-thick AGN with large Fe K$\alpha$ EWs that the sources with the most extreme EWs had evidence of on-going star formation within the centers of their host galaxies. Recently, \citet{boorman} noted that IC 3639 has an Fe K$\alpha$ EW above 2 keV and an X-ray spectrum that requires the \textsc{apec} component to fit the soft X-ray emission, with estimated X-ray and far-infrared SFRs above 10 M$_{\sun}$ yr$^{-1}$, consistent with starburst galaxies.

Our analysis on NGC 4968 further supports the idea that extreme Fe K$\alpha$ EWs can be an indicator that circumnuclear star formation may be present.  As shown by our imagining analysis, the soft emission, associated with star formation, extends to physical scales of $\sim$900 pc, such that the SFR density is several M$_{\sun}$ yr$^{-1}$ kpc$^{-2}$. Though this SFR is not prodigious, it seems clear we are observing X-ray emission from recent star formation processes within the central kpc of the galaxy. This result is consistent with a symbiotic relationship between starbursts and the obscuring torus: feedback from supernovae and stellar winds can puff up the torus \citep[e.g.,][]{wada2002,wada2009,schartmann}. If the torus is sufficiently inflated such that the covering factor approaches unity, this obscuring medium will preferentially attenuate the total continum with respect to the line photons, which has the effect of boosting the EW. The gas associated with star formation can perhaps also act as this obscuring material \citep[e.g.,][]{thompson,ballantyne} and ultimately fuel the black hole \citep{hobbs}, though the connection between nuclear star formation and AGN obscuration or black hole fueling depend on a number of unknown variables (e.g., gas fraction, outer radius of accretion disk, Mach number of accretion disk).

However, starburst/AGN composites may not always be systems that harbor extreme Fe K$\alpha$ EWs. For instance, Arp 220 is a well known starburst galaxy hosting an AGN \citep[e.g.,][]{ptak}, yet it lacks a strong neutral Fe K$\alpha$ line \citep{teng}.\footnote{Though an ionized Fe line is detected at $\sim$6.7 keV, this can only be formed in warm, optically thin matter, distinct from neutral gas with a high column density attributed to the putative torus.} Though AGN in starburst galaxies may not always be completely cocooned inside gas with a high column density, providing the requisite conditions to produce extreme neutral Fe K$\alpha$ EWs, AGN with these prominent features are prime candidates to search for signatures of on-going star formation.

\section{Conclusions}
We presented the {\it Chandra} imaging and spectral analysis of the Seyfert 2 galaxy NGC 4968. Capitalizing on {\it Chandra}'s arcsecond resolution, we investigated the spatially resolved X-ray emission, finding:
\begin{itemize}
\item The hard (2-10 keV) X-ray emission is extended on scales of $\sim$500 pc, coincident with the narrow line region as mapped by [OIII] 5007 \AA\ emission detected by the {\it Hubble Space Telescope} \citep{schmitt03}.
  
\item The soft (0.5-2 keV) X-ray emission is significantly extended to scales of about a kiloparsec, and is more extended than, and spatially distinct from, both the narrow line region and a previously detected elongated radio structure \citep{schmitt01}.
\end{itemize}

The most striking feature of NGC 4968 comes from its X-ray spectrum, which reveals a prominent Fe K$\alpha$ EW ($>$2 keV) as reported in \citet{guainazzi} and \citet{me2011}. From the {\it Chandra} spectrum, we measure an EW of 2.5$^{+2.6}_{-1.0}$ keV. This extreme value can only be achieved if the continuum is suppressed by Compton-thick levels of obscuration, and is more likely to occur in a geometry where the central engine is completely enshrouded. In that case, the differential extinction of the total continuum (emanating from the center of the obscuring medium, and including the reflected and scattered continua) compared with the fluorescent line emission (produced throughout the medium) is the greatest, boosting the EW.

We use the MYTorus \citep{mytorus} and the spherical absorption models of \citet{brightman}, which are physically motivated models that self-consistently treat the transmitted, reflected, and fluorescent line emission, to fit the X-ray spectra of NGC 4968. From the spectral analysis, we learn:
\begin{itemize}
\item The obscuration is at least near Compton-thick levels and is likely more extreme:
  \begin{itemize}
  \item The MYTorus fit to the spectrum returns $N_{\rm H}$ = 1.35$^{+0.22}_{-0.22} \times 10^{24}$ cm$^{-2}$, while the spherical absorption model measures an $N_{\rm H}$ of $7.64^{+8.57}_{-2.37} \times 10^{24}$ cm$^{-2}$.
  \item The instrinsic X-ray luminosity from the spherical absorption fit ($L_{\rm 2-10keV, int}$ = $7.8^{+0.7}_{-5.4} \times 10^{42}$ erg s$^{-1}$), about 2 orders of magnitude higher than that which is observed, is consistent with what would be predicted based on the MIR - X-ray correlation \citep{asmus,gandhi}.
  \end{itemize}

  A future significant high energy (i.e., $>$10 keV) observation from e.g., {\it NuSTAR} should differentiate between a toroidal and spherical absorption model. These data would provide the most accurate constraints on the column density and intrinsic X-ray luminosity.  

\item A thermal model (\textsc{apec}) is needed to adequately fit the soft X-ray spectrum, indicating the presence of hot gas with temperature kT $\sim$ 0.7 keV around the AGN. Such emission is likely linked with on-going star-formation.

\item We separately investigated the X-ray spectrum from the nuclear ($<$400 pc) and extended (400 pc - 1.3 kpc) regions:
  \begin{itemize}
  \item {\it Nuclear spectrum}: Thermal X-ray emission is present at the smallest scales we are able to resolve, with a temperature of kT $\sim$ 0.75 keV.
  \item {\it Extended emission spectrum}: Within the extended region, both thermal and non-thermal emission are present and are ascribed to star formation and unresolved X-ray binaries and/or scattered AGN photons, respectively. The temperature of the gas (kT $\sim$ 0.33 keV) is lower than that within the nuclear region.
  \end{itemize}

\end{itemize}

We derived the SFR based on the thermal X-ray luminosity ($L_{\rm thermal} \sim$ 6 $\times 10^{39}$ erg s$^{-1}$) and the template presented in \citet{me2012}, finding a SFR of $\sim$2.6-4 M$_{\sun}$ yr$^{-1}$, consistent with the far-infrared SFR (based on the 60$\mu$m and 100$\mu$m fluxes) of 2.4 M$_{\sun}$ yr$^{-1}$. Though this rate is not as prodigious as most starburst galaxies, we can conclude that star formation is on-going in the central $\sim$kpc of this galaxy.

It is interesting that AGN with extreme Fe K$\alpha$ EWs appear to reside in galaxies with circumnuclear star-formation \citep[e.g.,][]{levenson,boorman}. This association may be a natural consequence of nuclear star formation, where gas completely envelops the AGN \citep[perhaps due to stellar feedback inflating the obscuring medium and increasing its covering factor to unity;][]{wada2002,wada2009,schartmann}, providing the conditions necessary to produce large EW values. Though a global trend between star formation activity and X-ray obscuration is not always apparent \citep[see, e.g.,][]{me2011}, gas from star formation may contribute to the column density in these systems \citep{thompson,ballantyne}. We  emphasize that the converse is not necessarily true: not all starburst/AGN composities have prominent Fe K$\alpha$ EWs. However, AGN with large Fe K$\alpha$ EW values are promising candidates to look for clues of on-going, circumnuclear star formation. Such a search may be especially relevant at higher redshift where the star formation rates and central gas masses are much higher than in the present day Universe. 

We close by noting that though the Fe K$\alpha$ emission line is the most prominent feature in the {\it Chandra} spectrum, the line is spectrally unresolved. Follow up high resolution grating spectroscopy, or microcalorimeter observations, would spectrally resolve resolve this feature and determine whether it originates within the standard broad line region or putative torus \citep[e.g.,][]{yaqoob01,yaqoob03,shu}.

\acknowledgements
We thank the referee for a thorough reading of this manuscript and for providing constructive comments. Support for this work was provided by the National Aeronautics and Space Administration through Chandra Award Number GO5-16112X issued by the Smithsonian Astrophysical Observatory for and on behalf of the National Aeronautics Space Administration under contract NAS8-03060. SML is supported by an appointment to the NASA Postdoctoral Program at the NASA Goddard Space Flight Center, administered by Universities Space Research Association under contract with NASA. NAL is supported by the Gemini Observatory, which is operated by the Association of Universities for Research in Astronomy, Inc., on behalf of the international Gemini partnership of Argentina, Brazil, Canada, Chile, and the United States of America. PB and PG thank STFC for support (trant reference ST/J003697/2).

The scientific results reported in this article are based to a significant degree on observations made by the {\it Chandra} X-ray Observatory. This research has made use of software provided by the {\it Chandra} X-ray Center (CXC) in the application packages CIAO, ChIPS, and Sherpa. 

\dataset [ADS/Sa.CXO#obs/17126] {Chandra ObsId 17126}\\
{\it Facilities:} \facility{CXO}

\end{document}